\documentclass[]{aa501}

\usepackage{graphicx}

\begin{document}            

\title{Modeling the spectral energy distribution of galaxies. II.
Disk opacity and star formation in 5 edge-on spirals}

\author{ Angelos Misiriotis  \inst{1,2}
\and Cristina C. Popescu \inst{3,4,5,}\thanks{Research Associate at 
The Astronomical Institute of the Romanian Academy, Str. Cu\c titul de Argint 
5, 75212 Bucharest, Romania}
\and Richard Tuffs \inst {3}
\and Nikolaos D. Kylafis \inst{1,6}
}

\offprints{angmis@physics.uoc.gr }
\institute{ University of Crete, Physics Department, P.O. Box 2208, 710 03
Heraklion, Crete, Greece
\and Observatoire de Marseille , 2 place Le Verrier 13248 Marseille Cedex
4, France
\and Max Planck Institut f\"ur Kernphysik, Saupfercheckweg 1, D--69117
Heidelberg, Germany
\and The Observatories of the Carnegie Institution of Washington,
813 Santa Barbara Street, Pasadena, California 91101, U.S.A.
\and Max Planck Institut f\"ur Astronomie, K\"onigstuhl 17, D--69117 
Heidelberg, Germany
\and Foundation for Research and Technology-Hellas, P.O. Box 1527, 711 10 
Heraklion, Crete, Greece
}

\date{Received; accepted}

\abstract{
Using tools previously described and applied to the prototype galaxy NGC~891, 
we model the optical to far-infrared spectral energy distributions (SED)
of four additional edge-on spiral galaxies, namely NGC~5907, NGC~4013, UGC~1082 
and UGC~2048.  Comparing the model predictions with IRAS and, where available, 
sub-millimeter and millimeter observations, we determine the 
respective roles of the old and young stellar populations in grain heating. 
In all cases,
the young population dominates, with the contribution of the old stellar
population being at most 40\%, as previously found for NGC~891. 
After normalization to the disk area, the massive star-formation rate (SFR)
derived using our SED modeling technique,
which is primarily sensitive to the non-ionizing ultraviolet output 
from the young stellar population, lies in the range 
$7\times10^{-4}-2\times10^{-2}\,{\rm M}_{\sun}{\rm yr}^{-1}{\rm kpc}^{-2}$. 
This is consistent with normalized SFRs 
derived for face-on galaxies of comparable surface gas densities 
from H$_{\alpha}$ observations. Though the most active star-forming galaxy 
of the five in absolute terms, NGC~891 is not an exceptional system
in terms of its surface density in SFR.
\keywords{dust, extinction -- galaxies: spiral -- galaxies: stellar 
content -- galaxies: ISM -- infrared: galaxies -- submillimeter }
}

\authorrunning{Misiriotis et al.}
\titlerunning{Modeling the spectral energy distribution of galaxies. II.}

\maketitle

\section{Introduction}

Optical and Far-Infrared (FIR)/sub-millimeter (submm)
data from galaxies contain complementary 
information about the distribution of stars and dust,
from which intrinsic quantities of interest - the star-formation rate (SFR) and 
star-formation history - can in principle be extracted. This is especially 
relevant to systems having intermediate optical depths to starlight, or
to inhomogeneous systems with optically thin and thick components.
Many, perhaps most, of normal (non-starburst) gas-rich galaxies
in the local universe may fall into these categories.

On the one hand, optical data probes the colour and spatial 
distribution (after correction for extinction) of the photospheric
emission along sufficiently transparent lines of sight. This is
particularly useful to investigate older, redder stellar populations
in galaxian disks with scale heights larger than that of the dust. 
A radiation transfer modeling technique that can be applied to edge-on 
systems, where the scale height of the stars and dust extinction
can be directly constrained, was introduced by Kylafis \& Bahcall
(\cite{kylafis}) and subsequently applied on several  galaxies 
(Xilouris et al. \cite{xilouris1}; \cite{xilouris2}; \cite{xilouris3},
hereafter referred to as X97, X98, X99 respectively).  Radiative
transfer codes in combination with observations of nearby edge-on galaxies were
also used by Ohta \& Kodaira (\cite{ohta}) and Kuchinski et al. (\cite{kuchinski}). 

On the other hand, grains act as test particles probing the strength 
and colour of ultraviolet (UV)/optical interstellar radiation fields. 
This constitutes
an entirely complementary constraint to studies of photospheric emission.
In the FIR, grains are moreover detectable over the full range of optical
depths present in a galaxy. At least part of this regime is  inaccessible
to direct probes of starlight, especially at shorter wavelengths, even 
for face-on systems. This particularly applies to light from young stars
located in, or close by, the dust clouds from which they formed, since 
a certain fraction of the light is locally absorbed. Furthermore, 
there is at least a possibility that most of the remaining UV and
even blue light from young stars that can escape into the disk might
be absorbed by diffuse dust there. Observations from IRAS, IRAM (e.g. 
Neininger et al. \cite{neininger}; Dumke et al. \cite{dumke}), 
SCUBA (e.g. Alton et al. \cite{alton}; 
Israel et al. \cite{israel}; Dunne \cite{dunne}; Bianchi et
al. \cite{bianchi1}) 
and ISO (e.g. Haas et al. \cite{haas}; Stickel et al. \cite{stickel}) 
can provide information on the quantity and spatial
distribution of the dust within spiral galaxies.

A combined analysis of the whole UV-optical\-/FIR\-/submm\-/mm 
spectral energy distribution (SED) of galaxies seems to
be a promising way to constrain the problem. Radiative transfer codes for an
assumed ``sandwich'' configuration of dust and stars were applied by Xu \& Buat
(\cite{xu1}) and Xu \& Helou (\cite{xu2}) to account for the energy budget over the whole
spectral range. They considered in detail the relative contribution of the
non-ionizing UV photons and the optical photons in heating the grains.
However these calculations did not incorporate
a model for the dust grain emission, nor for radial
variations in the absorbed radiation in the disk, and
therefore could not account for the exact shape of the
FIR SED. Recently, there have been several
works modeling the SED of galaxies from UV to submm (Silva et al. \cite{silva};
Devriendt et al. 1999), by applying photometric and/or
spectrophotometric and chemical evolution models of galaxies. While these 
models are adequate in describing the general shape
of the volume-integrated  SED, they make use of many
free parameters and of a simplified geometry. 

Bianchi et al. (\cite{bianchi2}) attempted to model NGC~6946 from the UV to FIR
using a 3D Monte Carlo radiative-transfer code for a simplified
geometry of emitters (a single stellar disk). They concluded that the
total FIR output is consistent with an optically thick solution. However,
their model did not consider a grain size distribution for grains, 
stochastic emission of small grains, and
the contribution of localized sources within star-forming complexes.
This resulted in a poor fit of the FIR SED and a failure to reproduce
the IRAS flux densities.

Another work on modeling the UV to submm emission was presented by 
Efstathiou et al. (\cite{efstathiou}) for star burst galaxies which were treated as
an ensemble of optically thick giant molecular clouds centrally
illuminated by recently formed stars. This modeling technique
successfully reproduced the observed SED of M82 and NGC~6090. Such a
technique obviously cannot be applied to normal ``quiescent'' disk galaxies
dominated by emission from the diffuse interstellar radiation field.

In Popescu et al. (\cite{popescu}, hereafter referred to as Paper I)
we gave a detailed description of a new tool for a combined analysis
of the optical-FIR/submm SED of galaxies.  This tool included
solving the radiative-transfer problem for a realistic distribution of
absorbers and emitters and by considering realistic models for dust, 
taking into account the grain size distribution and stochastic heating
of small grains as well as the contribution of HII regions.
We applied this tool to the edge-on system NGC~891.

In short, we used the intrinsic
spatial distribution of dust and stars (in the B, V, I, J, K bands) derived for
this galaxy by X99 as a constraint on 
the old stellar population and part of the dust distribution.
In order to fully explain the optical-FIR-submm SED, it was found
necessary to add to the components of the X99's solution both a
young stellar population (to correct for a shortfall in
FIR brightness in terms of re-radiated non-ionizing UV)
and more dust mass (to correct for a shortfall in submm brightness). 
Our solution explained both the observed SED in the FIR and submm, as
well as the observed radial profile at 850\,${\mu}$m. We found that the dust is
predominanly heated by the young stellar population.

Although NGC~891 was chosen for study in Paper I as the prototypical and most
extensively observed edge-on galaxy, it exhibits some extreme features,
raising the possibility that it might not in fact be representative of
a ``typical'' spiral. In this paper we extend our SED modeling 
technique to four additional edge-on systems - NGC~5907, NGC~4013, UGC~1082 and 
UGC~2048 - with the aim of examining whether the features of the 
solution we obtained for NGC~891 might be more generally applicable. 
Radiation transfer solutions accounting for the appearance of 
these four galaxies in the optical bands have been obtained 
by X99, X98 and X97.  We use these solutions to constrain our model for the 
optical-FIR SEDs in the same fashion as for NGC~891.

The paper is arranged as follows:
In Sect. 2 we overview the model and its application to the objects
in the present work.
In Sect. 3 we present predictions for the FIR SED of
all five galaxies studied. 
These are obtained using our ``standard model'', i.e. a model
with radiation fields derived from the radiation transfer solution for the
optical appearance supplemented by a hidden population of young stars 
embedded in HII regions, but with no additional dust.
In Sect. 4 we describe and discuss the SED of one of our four galaxies, 
namely NGC~5907, in terms of the more complex ``two-dust-disk model'' and 
compare this with the corresponding solution obtained for NGC~891 in Paper I.
In Sect. 5 we discuss the dependence of the derived star-formation
rates on the assumptions of the model, in particular the assumed geometries 
for the dust and stars. We show that after normalization to disk area, 
NGC~891 has comparable star-formation characteristics to the other four 
objects. We also compare the disk-area normalized massive 
star-formation rate in the five edge-on galaxies, obtained from the 
optical-FIR SED analysis, with the same quantity derived from H$_{\alpha}$
measurements of statistical samples of face-on galaxies given in the
literature.  In Sect. 6 we give a summary of our work.

\section{Model calculations for the five galaxies}
A full description of our model and its application
to NGC~891 can be found in Paper I.
Here we give an overview of the basic idea and its specific application
to the four galaxies.

To determine the propagation of light in a galaxy we take
both emitters and absorbers to be smoothly arranged as a
superposition of simple cylindrically symmetric geometrical distributions.
The assumption of an axisymmetric model and the neglect of the spiral
structure does not seem to introduce a systematic error on the estimate of
the overall opacity, at least when a statistical sample of edge-on galaxies is
considered (Misiriotis et al. \cite{misiriotis}).  The emitters are divided into a young 
(predominantly UV-emitting) stellar population, distributed in a disk with 
a small scale height, 
an old (predominantly optical-NIR emitting) 
population, distributed in a disk with a larger scale height and a bulge. 
The absorbers are arranged in ``young'' and ``old dust disks'', 
associated with the old and young stellar populations. The ``old dust 
disk'' has a larger scale height than the ``young dust disk'', though not
so large as the old stellar disk. Both the stellar- and dust-disk 
distributions are specified as exponential density distributions in radius
and height, while the stellar bulge is described by a de Vaucouleurs law. 
All radial distributions have a common truncation at three scale lengths of the
``old dust disk''. 

The first step in the calculation of the optical-FIR SED is to determine the
geometric and amplitude parameters for the old stellar disk and bulge 
independently in each band pass ranging from the B band to 
the NIR for a common ``old dust disk'' (whose geometry and opacity is also 
determined). This can be achieved by a radiation transfer analysis matching
as closely as possible the predicted and observed brightness
distributions, following
Kylafis \& Bahcall (\cite{kylafis}) and including
inclination as a free parameter.
For the five galaxies discussed in this work we
adopt the solutions given in X97 and X99 on the basis of
observations in B, V and I, and in the case of NGC~891 also in J and K bands.

This analysis of the optical-NIR emission also 
yields the extinction coefficient as a function 
of wavelength for the ``old dust disk''. For all five galaxies 
studied, the measured extinction law was found to be close to the 
predicted law derived from the dust model used in the 
calculation of the IR emission. This corresponds to 
the graphite-silicate mix and dust size 
distribution determined for the Galaxy by Mathis et al. (\cite{mathis}) 
(see Eqs. [2] and [3] of Paper I) with optical constants taken from
Laor \& Draine (\cite{laor}). When calculating the IR emission we take
grains in the ``young dust disk'', if present, 
to have the same size distribution and optical properties as those in the 
``old dust disk''.

To this fixed basis for the old stellar population and associated
dust disk, the model allows the addition of a young, UV-emitting 
stellar population and, optionally, associated dust. The additional stars 
and dust are taken to be distributed in a common exponential disk with a 
scalelength equal to that of the already determined intrinsic B-band 
population, and a scale height fixed at 90\,pc (as for the Milky Way). 
Since this scale height is typically a factor of several times smaller 
than the scale heights derived by X97 and X99 for the ``old dust disk'', 
this constitutes the simplest assumption for the distribution of these 
extra components, which hide UV-optical indicators of the  
young stellar population and associated additional 
opacity. As emphasized in Paper I, a clumpy distribution
for the dust associated with the young stellar population is also
possible, though very difficult to calculate, and, due to the
lack of high angular resolution FIR images of spiral galaxies, at
present not directly verifiable.
The luminosity of the young stars is a free parameter of the 
model, which we express as a recent star-formation rate SFR, based 
on the population synthesis models of Bruzual \& Charlot (\cite{bruzual}) for 
$Z=Z_{\odot}$, a Salpeter IMF with a cut-off mass of 
$100\,{\rm M}_{\odot}$ and $SFR=(1/\tau) {\rm exp}(-t/\tau)$ with ${\tau}=5\,$Gyr.

The model also incorporates a parameter $F$ representing
the fraction of the non-ionizing UV luminosity from the young 
stellar population, which is locally absorbed in star-formation 
regions by optically thick fragments of their parent molecular clouds.
This fraction of the luminosity is considered to be re-radiated 
in the FIR according to a spectral template matching the 
observed spectrum of galactic HII regions. This template 
generally has a warmer 60/100\,$\mu$m colour than that calculated 
for the diffuse disk.

Given the intrinsic distributions of stellar emissivities in the UV,
optical and NIR, and the distribution of absorbers in the ``old'' 
and (if present) the ``young dust disks'', we then proceed with a radiative
transfer calculation to determine the UV-NIR energy density of the 
radiation field throughout the galaxy. In the absence of solutions in
the J and K bands, we extrapolate the energy density from the B, V, I 
bands assuming a black-body spectral distribution 
and a colour temperature of 4000 K. In calculating the heating of grains
placed in the resulting radiation field, we include an explicit treatment of 
stochastic emission as detailed in Paper I.  The IR-submm emission from 
grains is then calculated for a grid of positions in the galaxy. 
Subsequently we integrate over the entire galaxy to obtain the IR-submm 
SED of the diffuse disk emission.
Prior to comparison with observed FIR-submm SEDs, the spectral template 
for the HII regions, scaled according to the value of $F$, must be 
added to this calculated spectral distribution of diffuse FIR emission.

Due to the precise constraints on the distribution of stellar emissivity in 
the optical-near infrared (NIR) and the distribution of dust in the 
``old dust disk'' yielded by the radiation transfer 
analysis of the highly resolved optical-NIR images, coupled with the
simple assumptions for the distribution of the young stellar 
population and associated dust, our model has a maximum of three free 
parameters from which the FIR-submm SED can be fully determined. These are 
the SFR for massive stars, the parameter $F$ representing the fraction of the 
non-ionizing UV luminosity from massive stars which is locally absorbed,
and the mass of dust in the ``young dust disk''. 
In general terms, the last parameter is principally constrained by 
the submm emission, the non-ionizing UV luminosity by the bolometric
FIR-submm output and the  factor $F$
(in the absence of high resolution images) by the FIR colour. 
In the analysis of the five galaxies in subsequent sections, we refer to 
the following variants of the model:

\begin{enumerate}

\item
{The {\it simple model}: The simplest possible
model takes into account only the information from the optical observations,
i.e. the dust and old stellar distribution in the old disk without 
massive UV-emitting stars. The sole purpose of the ``simple model'' is to 
demonstrate that the old stellar population can not adequately 
heat the dust to account for the observed FIR luminosity.}
\item
{The {\it standard model}: This is the ``simple model'' 
supplemented by a disk population of UV-emitting stars 
without associated dust, but with localised FIR emission.
The free parameters are the SFR and the  factor $F$.}
\item
{The {\it two-dust-disk model}: This is the ``standard model'' 
supplemented by the presence of dust in the young stellar disk. 
The free parameters are the SFR, the factor $F$, and the dust mass in the
``young dust disk''.}

\end{enumerate}

The predictions of the model calculations are compared with available FIR,
submm and mm data from the literature. In Table~\ref{tab1} we summarise the data used for
comparison in the present work. We adopted $20\%$ errors in the IRAS and
IRAM data.  For the SCUBA data we adopted
the errors quoted by Alton et al. (\cite{alton}). 
In Table~\ref{tab1} we also give some basic information on the
properties of the sample galaxies. One factor in the choice of
these galaxies for the radiation transfer analysis by X97 and X99
was their prominent dust lanes.  

\begin{table*}
\caption{Properties of the galaxies used in the modeling of the
FIR-submm SEDs. The dust mass is that quoted in X97 and X99 
(corresponding to the ``old dust disk'' in this paper). 
The submm fluxes are taken from
$^1$Alton et al. (\cite{alton}), 
$^2$Young et al. (\cite{young}), $^3$Soifer et al. (\cite{soifer}), $^4$Moshir et al. (\cite{moshir}).
The mm fluxes from $^5$Dumke et al. (\cite{dumke}), $^6$Guelin et al. (\cite{guelin}.}
\begin{tabular}{clrrcccccc}
\hline
Galaxy &Hubble&distance&dust mass              &$F_{60}$        &$F_{100}$   
&$F_{450}$&$F_{850}$&$F_{1200}$&$F_{1300}$\\
       &type  &(Mpc)   &($M_{\sun}\times 10^7$)& (Jy)           & (Jy)       &  
(Jy)   &  (Jy)  &  (Jy)     & (Jy)     \\
\hline

NGC~891 & Sb  &9.5    &  5.6                  &$50.5^1$        &$126^1$      
&$32^1   $&$4.62^1$ & -        &$0.85^6$ \\ 
NGC~5907& Sc  &11.0     &  1.5                  &$16.3^2$      &$55.9^2$& -      
 & -       &$0.54^5$  & -       \\ 
NGC~4013& Sbc &11.6   &  0.45                 &$7.0^3$         &$23.1^3$     & - 
      & -       & -        & -       \\ 
UGC~1082& Sb  &37.0     &  0.99                 &$1.6^4$         &$3.7^4$      & 
-       & -       & -        & -       \\ 
UGC~2048& Sb  &31.5   &  3.5                  &$1.7^4$         &$3.5^4$      & - 
      & -       & -        & -       \\ 
\hline
\end{tabular}

{\footnote{} The flux densities from this reference 
are integrated
in the range $\pm 225^{\prime\prime}$ along the major axis of
the galaxy.}
\label{tab1}
\end{table*}

\section{Results for the ``standard model''}

\begin{figure*}
\resizebox{\hsize}{!}{\includegraphics[width=17cm]{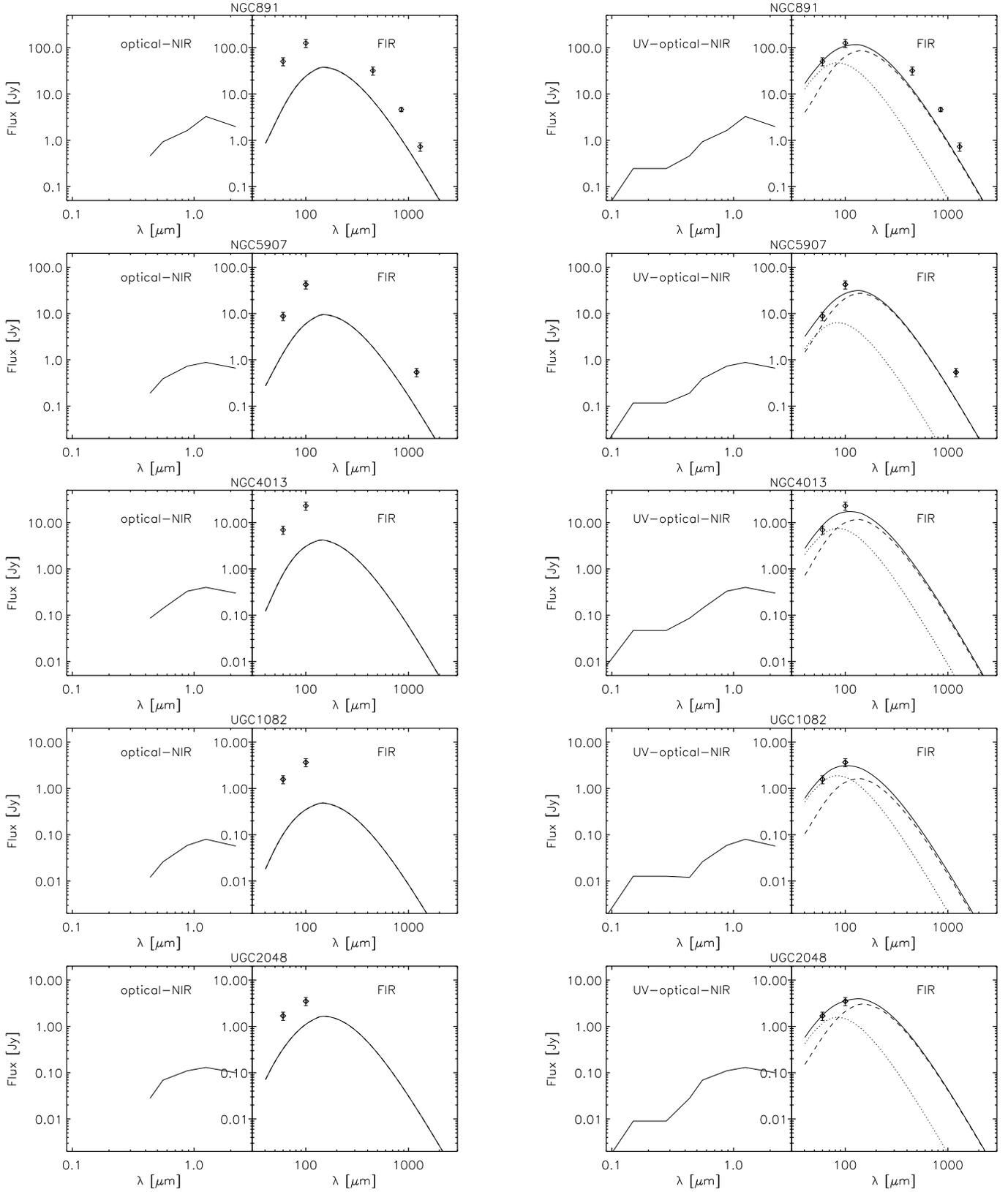}}    
\caption{The SED of the 5 galaxies. In the left column the SED is derived for
the ``simple model''. In the right column the SED is derived for the ``standard
model''. The predicted FIR emission from the diffuse disk is plotted with
dashed lines and the contribution of localised sources with dotted lines. The
corresponding SFR and factor $F$ for each galaxy are given in Table~\ref{tab2}. For 
NGC~891 and UGC~2048, the model corresponding to the value of $F$ which best 
fits the IRAS colour is plotted. In the cases of NGC~5907, NGC~4013 and 
UGC~1082,
the model corresponding to the maximum value of $F$ still consistent with the 
IRAS colour is plotted (see text).
}
\label{fig1}
\end{figure*}

To demonstrate the need for a population of massive stars to account for
the observed FIR luminosities in all five galaxies, we first calculated  
SEDs for the ``simple model'', in which the stellar and dust distributions 
are completely determined from the optical-NIR data as described in Sect. 2.
These are plotted in the left column of Fig.~\ref{fig1} In all cases, 
the predicted FIR flux densities in the 60 and 100\,$\mu$m IRAS bands 
fall short of the observed luminosities by factors of between 5 and 10,
showing that the heating of the dust from the old stellar
population cannot account for the FIR emission.
The results of the ``simple model'' are summarised in the second column
of Table~\ref{tab2} ($L_{\rm simple}$), where we give the FIR luminosity of the 
galaxies that can be attributed to the old stellar population.

Assuming that the optically determined dust content we used for each 
galaxy is exact, the ``standard model'' can account for the 
observed FIR luminosity by transforming non-ionizing UV radiation from 
massive stars into the FIR through a combination of local absorption
and absorption in the diffuse old disk, controlled by the factor $F$.
If all the young stellar luminosity is absorbed locally ($F=1$),
then the efficiency of the transformation from non-ionizing UV to the 
FIR is unity, as for starburst galaxies. In the total absence of 
local absorption ($F=0$) the global absorption efficiency is lower, 
with a value depending on the relative geometries of the diffuse dust and 
the massive stars. In principle, the factor $F$ can be determined from
the FIR colours.

We investigated the combinations of $F$ and SFR for which the ``standard 
model'' was consistent with the IRAS 60 and
100\,${\mu}$m measurements. For NGC~891 and UGC~2048, good agreement
between model and data could be found for values of $F$
well above zero. However, for the remaining galaxies,
the IRAS colours required $F$ values of zero or close to zero. This
would imply high current star-formation rates and an amplitude
in the UV stellar emissivity which would not join up smoothly
with the empirically determined B band stellar emissivities.
Consequently, we have plotted in Fig.~\ref{fig1} the model prediction
corresponding to the maximum value of $F$ still consistent with the
IRAS colours. This also gives a monotonically increasing SED
for the UV-optical stellar emissivities. Corresponding
FIR bolometric luminosities ($L_{\rm standard}$), SFRs and factors  $F$
are given in Table~\ref{tab2}, columns 3, 4, 5 respectively. 
We emphasise however, that in 
principle one could have a non-steady SFR, so that the tabulated SFR 
are at the same time lower limits in SFR and upper limits for the factor $F$.
Therefore we can discuss the results for the ``standard model'' also in
terms of a range of possible values for $F$ and SFR, which will be compatible
with the range in the IRAS errors. 

We remark that a comparison of the values in the second and third
column of Table~\ref{tab2} reveals that, according to the standard model,
only 10\% to 30\%  of the dust emission is powered by the old 
stellar population. 

In the above we have made the assumption that the ``standard model'' is a 
sufficiently realistic
model of the galaxies. Some evidence that this may not be the
case is apparent from Fig.~\ref{fig1} (right column), where all the model spectra
peak at around  120\,$\mu$m,
well shortward of the typical peak near 150\,$\mu$m seen in spiral
galaxies. This may be an indication that the overall optical
depth of the galaxies is underestimated by the ``standard model'',
indicating the presence of a further dust component in the plane of each 
galaxy. This component may either be smoothly 
distributed or be in the form of optically thick clumps or some combination
of these.
For the two galaxies that submm and/or mm measurements are available
(NGC~891 and NGC~5907), this is explicitly confirmed by the undershoot
of the ``standard model'' prediction in the submm and mm regime.
In Paper I, we used the available submm fluxes to quantify the missing 
dust and introduced the ``two-dust-disk model'', which successfully reproduced
the spectrum and the spatial distribution of the FIR emission in
NGC~891. 
In the next section we will apply the ``two-dust-disk model'' to NGC~5907 to 
determine the total dust content of this galaxy and examine 
the effect of this more realistic case on the derived SFRs.

\begin{table}
\caption{Results for the ``standard model''.
$L_{\rm simple}$ denotes the bolometric re-radiated 
luminosity when only the old stellar population (B, V, I, J, K bands) 
is taken into account (i.e. for the ``simple model''). 
$L_{\rm standard}$ denotes the bolometric re-radiated luminosity obtained
by integrating the solution for the ``standard model'' as plotted in Fig.~\ref{fig1}. 
Columns 4 and 5 contain the corresponding SFRs and factors $F$. }

\begin{tabular}{ccccc}
\hline
Galaxy & $L_{\rm simple}$ & $L_{\rm standard}$ & SFR & $F$ \\
       & W$\times10^{35}$ & W$\times10^{35}$ &  M$_{\sun} {\rm yr}^{-1}$  & \\
\hline
NGC~891 & 10.8  & 48.1      &      3.48   &   0.28     \\
NGC~5907&  3.7  & 15.5      &      2.17   &   0.08     \\
NGC~4013&  2.0  & 11.0      &      1.23   &   0.19     \\
UGC~1082&  2.3  & 21.3      &      2.59   &   0.23     \\
UGC~2048&  5.7  & 18.3      &      1.36   &   0.27     \\
\hline
\end{tabular}
\label{tab2}
\end{table}

\section{Application of the ``two-dust-disk model'' to NGC~5907}

In Paper I 
we proved that in NGC~891 the amount of dust
(on the basis of the grain optical constants adopted) quoted by X99 
is underestimated by a factor of two, and that 
the addition of a second dust disk, that follows the spatial
distribution of the young stellar population, reproduces
very well the observed FIR luminosity and the radial profiles
observed by Alton et al. (\cite{alton}). 
For NGC~5907, the measurement at 1200\,$\mu$m by Dumke et al. (\cite{dumke})
can be used to constrain the quantity of dust in the second disk.

Inspection of Fig.~\ref{fig1} (right column)
shows that the ``standard model'' solution for $F=0.08$,
which is in agreement with the IRAS points, crassly undershoots the 
1200\,$\mu$m point.
As for NGC~891, the only way to account for both the IRAS and the 1200\,$\mu$m 
observations is to include more dust in our model, as simply
increasing the star-formation rate would provide too much luminosity
in the IRAS range.

\begin {figure}[!t]                                                           
\resizebox{\hsize}{!}{\includegraphics{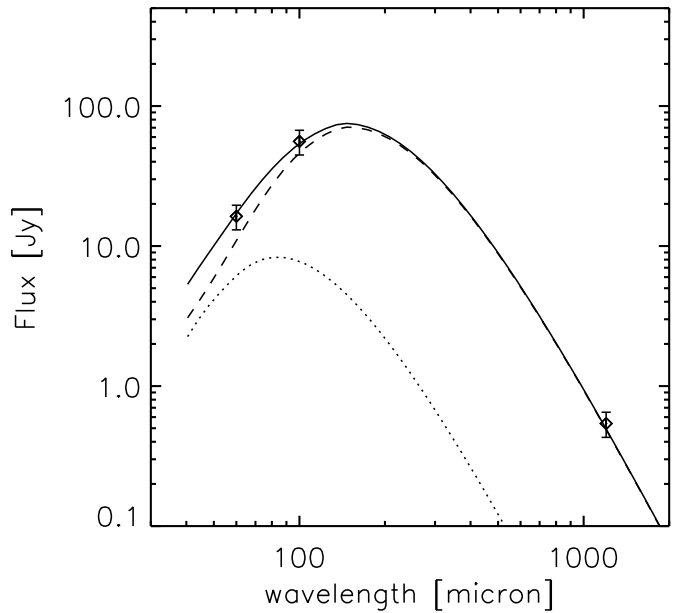}}    
\caption{SED of the ``two-dust-disk model'' for NGC~5907 for $F=0.10$
and SFR $= 2.2\,{\rm M}_{\odot} {\rm yr}^{-1}$,
overlaid with data points given in Table~\ref{tab1}. 
The predicted FIR emission from the
diffuse disk is plotted with a dashed line and the contribution of localized
sources with a dotted line.} 
\label{fig2}
\end{figure}

The second-dust-disk's spatial distribution was constrained
to follow that of the young stellar population, which
is taken to be exponential with scalelength and scaleheight 
$h_{\rm d}=5020$\,pc and $z_{\rm d}=90$\,pc respectively (see Sect. 2). To find the
solution, the total amount of dust in the second dust disk
was varied jointly with the SFR until the FIR-submm SED of our model 
fitted all three measurements. In Fig.~\ref{fig2} we present the resulting 
spectrum overlaid with the data. This spectrum corresponds to 
a SFR$= 2.2\,{\rm M}_{\sun} {\rm yr}^{-1}$ and $F=0.10$. The total
dust mass in the second dust disk is $4.5 \times 10^7\,{\rm M}_{\sun}$. 
By comparison, the dust mass quoted for the old disk of NGC~5907 by X99 
is $1.5 \times 10^7\,{\rm M}_{\sun}$. 

Thus, we have had to add 3 times more
dust over the quantity implied by the observed dust lane to obtain agreement
with the data
on the basis of the ``two-dust-disk model''. The derived value of $F$ is still 
very low, but the spectrum of the diffuse component is too warm 
(NGC~5907 is more optically thin than NGC~891) to allow the addition of a 
warm component related to HII regions. While we could
have produced a colder spectrum for the diffuse emission, allowing some 
room for HII regions, this would have been at the cost of adding yet more dust
in the second disk. A larger quantity of dust in the second disk
does not seem very likely, as the total dust mass for the fitted $F=0.10$ 
case already implies a gas-to-dust ratio of 130 by mass 
(instead of 520 according to X99). As a comparison, the gas-to-dust ratio in
the Milky Way is 133.

The total IR-submm re-radiated luminosity of NGC~5907, obtained by
integrating the $F=0.10$ ``two-dust-disk model'' SED, is
$50.5\times 10^{35}$\,W out of which $27.0\times 10^{35}$\,W
is attributed to heating from the young stellar population.
Thus, about $40\%$ of the 
dust emission is powered by the old stellar population. More dust in the 
``young dust disk'', a lower SFR and 
a higher factor $F$ would increase the disk 
opacity and thus the contribution of the old stellar population, but 
we excluded this solution due to the high dust-to-gas ratio. The $40\%$
contribution from the old stellar population represents
an upper limit, since a possible clumpiness in the dust
distribution would decrease the absorption efficiency in the optical
band.

On the basis of the model's inherent assumption, that
all the dust (in both disks) of the galaxy is distributed smoothly, 
the central face-on optical depth in the optical band 
is ${\tau}_{\rm v}=1.4$ for NGC~5907.

The major difference between
NGC~891 and NGC~5907, on the basis of the ``two-dust-disk model'', 
is that the spectrum of the former apparently allows for the existence of a 
larger contribution from HII regions. 
Presumably NGC~5907 must have HII regions, and
there is no obvious reason why the local properties of these HII
regions, in particular the probability of local absorption of the
UV radiation, should differ between the two systems. 

One potential
way out of this difficulty might be to suppose that the spectral
template for the HII regions used in our model underestimates the
associated submm emission. As stated in Paper I, we did not attempt 
to include potential cold dust emission components that might be expected 
from ``parent'' molecular clouds in juxtaposition to their ``offspring''
HII regions. These might contribute a significant fraction of the observed
submm emission, but be relatively faint in the IRAS range. 
If so, the dust content and the luminosity
of the diffuse disk would be lower for both galaxies, but more so
for the more optically thin diffuse disk of NGC~5907. This would 
make room for larger values of $F$ in the model fits, particularly 
for NGC~5907.

An alternative scenario, also discussed in Paper I,
would be the presence of truly quiescent cold dust clouds in the disk
not associated with HII regions. 
In that case, the difference between the
two galaxies might indicate a greater star-formation activity in the disk
of NGC~891, in the sense that a larger proportion of the cold dust
clouds in the disk had been triggered into massive star formation.

\section{Discussion}

Our results confirm that it is a common tendency for edge-on galaxies
to hide a significant fraction of their dust from optical extinction
studies. In two cases where sub- or near-mm data are available
(NGC~891, as analyzed in Paper I and NGC~5907 in this work),
we have modeled the FIR-submm SED assuming that the additional dust is 
distributed in a second diffuse disk associated with the young stellar 
population. We have found that the SEDs could be fitted in terms of 
total dust masses of respectively twice and four times the
masses inferred by X99 from the optical radiation transfer analysis
for NGC~891 and NGC~5907. The corresponding SFRs and local non-ionizing
UV absorption factors $F$ in the best fits were 
SFR $=3.8$\,M$_{\odot} {\rm yr}^{-1}$ and $F=0.22$ for NGC~891 and 
SFR $=2.2$\,M$_{\odot} {\rm yr}^{-1}$ and $F=0.10$ for NGC~5907.

The main uncertainty in the inferred SFRs is the value of the factor  $F$.
In principle, $F$ can be
determined from the colour of the FIR SED, especially 
the 60/100\,$\mu$m colour ratio. However, as
discussed in Sect. 4, the value of the factor $F$, which
can be fitted, in practice is constrained by the assumption of the   
``two-dust-disk model'' that all the additional dust needed to fit the submm
measurements is diffuse. In reality, the geometry of the additional dust 
could be, as emphasised in Paper I, in the form of 
optically thick clumps associated with the HII regions. 
In this case the inferred SFRs would, depending on the factor $F$, 
differ from those determined on the assumption of a purely diffuse 
dust distribution without local absorption. However, at present,
the available data does not allow us to quantify the clumpy 
component of UV-absorbing dust.

The SFRs and factors $F$ inferred from the 
``standard model'' for NGC~891 and NGC~5907 in Sect. 3 (where the 
dust content and distribution is determined from X99's optical analysis) 
hardly differs from the values determined from the ``two-dust-disk model''
applied to NGC~891 and NGC~5907 in Sect. 4. This is at first sight
surprising, as one would expect lower SFRs due to the higher opacity
of the diffuse disk in the ``two-dust-disk model''. However, the SFR
determined from the ``standard model'' was obtained by fitting the SEDs 
exclusively to the IRAS 60 and 100\,$\mu$m measurements common to the
sample of the 5 galaxies.
The reason that the SFRs determined from the ``standard model'' are so
similar to those derived from the ``two-dust-disk model'' is simply
that the additional submm luminosity arises from the extra dust (from
the second dust disk) rather than from additional SFR.
As remarked in Sect. 3, 
the rather warm SEDs for all the ``standard model'' fits suggest that 
additional dust is present in the disks of all 5 galaxies.

To statistically evaluate our results for SFR,
we make the assumption that the SFRs determined from the ``standard model'' 
fits to the IRAS 60 and 100\,$\mu$m measurements for our sample of 5 
galaxies give, 
as is the case of NGC~891 and NCG~5907, a reasonable estimate of the 
SFR which would have been obtained with an analysis based on the 
``two-dust-disk model''. We can then compare the SFR characteristics of the 5 
edge-on 
galaxies with the larger sample of 61 galaxies with inclinations less
than 75 degrees which were studied by Kennicutt (\cite{kennicutt}) on the basis of 
H$_{\alpha}$ measurements.

\begin {figure}[!t]                                                    
\resizebox{\hsize}{!}{\includegraphics{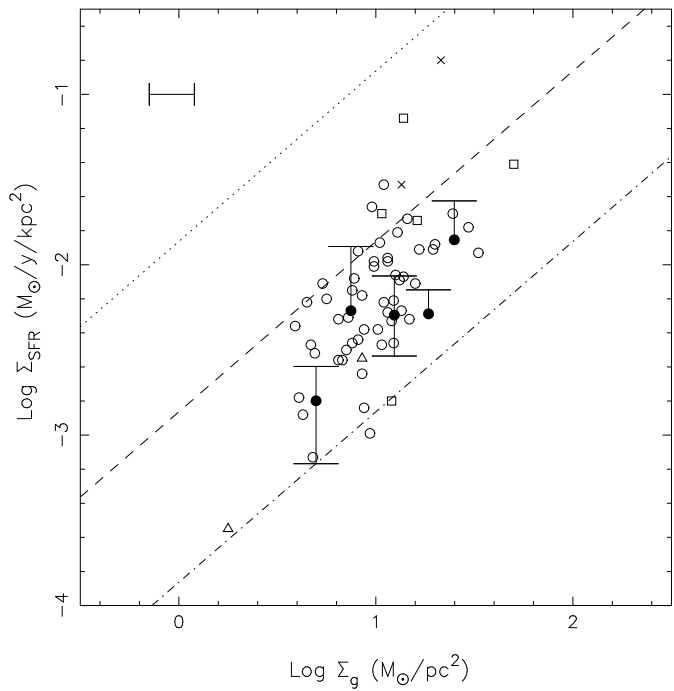}}    
\caption{Disk-averaged SFR surface density ($\Sigma_{\rm SFR}$) as a function 
of average gas surface density ($\Sigma_{\rm g}$) for our galaxy sample and for
the sample of Kennicutt (1998) of 61 normal disk galaxies with SFR 
determined from H$_{\alpha}$ measurements. 
The 5 galaxies from our sample are plotted as filled circles 
and the corresponding
$\Sigma_{\rm SFR}$, $\Sigma_{\rm SFR}^{\rm min}$, 
$\Sigma_{\rm SFR}^{\rm max}$, $\Sigma_{\rm g}$ and disk areas are listed 
in Table~\ref{tab3}.
The SFR surface densities were calculated by averaging the SFRs determined 
from the ``standard model'' (Sect. 3) over disks with an optically defined 
boundary
($R_{\rm o}$) taken to be 3 times the intrinsic radial scalelength $h_{\rm s}$ 
determined from the radiation  transfer modeling in the I band. 
The galaxies from the sample of Kennicutt are plotted as open circles 
(Sb,Sc,SBb,SBc), triangles (Sa), open squares  (Unknown/Not Available),
crosses (Irr).
The dotted, dashed and dot-dashed lines represent
star-formation efficiencies corresponding to consumptions of 100,
10 and 1~percent of the gas mass in 10$^{8}\,$yr.}
\label{fig3}
\end{figure}

Fig.~\ref{fig3} depicts the relation between the disk-averaged surface density in 
SFR ($\Sigma_{\rm SFR}$) determined from the ``standard model'' as a 
function of the average gas surface density ($\Sigma_{\rm g}$) for the 
5 galaxies. The plotted data is summarized in Table~\ref{tab3}. 
The $\Sigma_{\rm SFR}$ is calculated from the SFRs determined
from the ``standard model'' and listed in Table~\ref{tab2}. 
The full range in conceivable SFRs is given by the vertical error bars.  
The upper and lower limits for the SFRs are calculated such that the 
predicted SED is still consistent with the IRAS colours 
(within the 20$\%$ IRAS error bars). Lower error bars are not given when 
the plotted SFRs represent lower limits for the SFR (maximum values for the 
factors $F$, see discussion in Sect. 3). In these cases lower limits would be
possible only if we allowed for different sources of uncertainties, like
variations in the spectral shape of the template used for the
HII regions. However this is hard to quantify, and in the following we assume
that the errors of the SFR are given only by the uncertainties in
the IRAS data. 

The $\Sigma_{\rm g}$ were calculated from the gas masses taken from
Garc\'{\i}a-Burillo et al. (\cite{garcia}) and Rupen (\cite{rupen}) 
respectively for H$_2$ and HI in NGC~891,
from Dumke et al. (\cite{dumke}) for H$_2$ and HI in NGC~5907,
from Bottema (\cite{bottema}) and Gomez \& Garcia  (\cite{gomez}) for H$_2$ and HI in NGC~4013,
from Giovanelli \& Haynes (\cite{giovanelli}) for HI in UGC~1082
and from Huchtmeier \& Richter (\cite{huchtmeier}) for HI UGC~2048. In the last two cases, 
where only data for HI was available, we estimated the total gas mass
by doubling the HI gas mass.
The horizontal error bar
corresponds to the uncertainty in the gas masses for which we adopted an average
0.2 dex error. The surface area of the disk was calculated for
$R_{\rm o}=(3 \pm 0.5) h_{\rm d}$, where $h_{\rm d}$ is the intrinsic 
radial scalelength determined from the
radiation-transfer modeling in the I band.
In their analysis of surface photometry of
the outer regions of spiral disks, Pohlen et al. (\cite{pohlen}) show that
the disk boundaries are typically in this range.

The points for the 5 galaxies in Fig.~\ref{fig3} lie within the
area of the diagram occupied by the galaxies in the Kennicutt sample. The 
match is even better for those members of the Kennicutt sample with  
Hubble types Sb to Sc.  
This agreement is quite reassuring, bearing in
mind the several factors which could introduce a systematic
difference between the SFRs inferred for a sample of nearly face-on systems
from H$_{\alpha}$ measurements compared with the present technique for
edge-on systems based on an analysis of broad-band non-ionizing UV 
re-radiated in the FIR-submm range.

Firstly, the H$_{\alpha}$
analysis is sensitive to the most massive stars and in particular to the
assumed mass cut of 100\,M$_{\odot}$. Whereas the FIR-submm modeling also
assumes the same mass cut in the conversion of SFR to non-ionizing luminosity
(see Sect. 2 and Paper I), our model is less sensitive to this effect.

Secondly, whereas the H$_{\alpha}$ is sensitive to the star-formation
history of the last 10$^{7}$yr, our broad-band FIR-submm SED analysis samples 
approximately the last 10$^{8}$yr. Thus, our analysis is consistent
with the basic hypothesis (see Kennicutt \cite{kennicutt}) for ``normal'' spiral
galaxies of a steady star-formation activity. In principle, we could
extend our analysis based on our determinations of the intrinsic
populations and use the determined intrinsic colours
to determine more accurately the SFR history of the galaxies, though
this is beyond the scope of this paper.

The assumption of a steady-state star-formation rate is also broadly
consistent with the timescales for the exhaustion of the current gas supply
under the derived SFRs.
The dotted, dashed and dot-dashed lines in Fig.~\ref{fig3} represent
star-formation efficiencies corresponding to consumptions of 100,
10 and 1~percent of the gas mass in 10$^{8}$\,yr.
We note that in constructing Fig.~3, we have assumed that
the HI line emission, from which the gas masses for the 5
edge-on galaxies were in part determined, have not been
affected by self absorption in the 21cm line. 
The effect of self-absorption would be to 
shift the points for the 5 edge-on galaxies to the left of the
positions corresponding to their actual gas surface densities in Fig.~\ref{fig3}.

Thirdly, the SFRs derived from H$_{\alpha}$ were corrected by a single
factor for extinction, despite the varying orientations. As well as 
possibly affecting the vertical position of the galaxies on the plot, 
this may induce some scatter, especially if all the dust were diffusely
distributed. The systematic effect may be expressed in terms of the factor $F$:
an overestimation of the factor $F$ is equivalent to an overestimation of the
local extinction in the star-formation regions (statistically averaged 
over the population of HII regions in a disk). Thus, while moving to higher  
factors $F$ would move the points for the 5 galaxies towards lower SFRs, it
would have the opposite effect for the SFRs determined from the H$_{\alpha}$.

Lastly, we remark that NGC~891 does not appear as an exceptional system
compared with the other 4 galaxies in our sample (and with Kennicutt's [1998]
normal galaxy sample) on the basis of SFR normalized to disk area. 
Our work thus provides no evidence that this galaxy's exceptional layer of 
extraplanar H$_{\alpha}$-emitting diffuse ionizing gas 
(e.g. Hoopes et al. \cite{hoopes}) and surrounding 
X-ray-emitting hot gas (e.g. Bregman \& Houck \cite{bregman}) is attributable to
unusual star-formation activity.

\begin{table}
\caption{The average gas surface density (${\Sigma}_{\rm g}$) and
disk-averaged SFR surface density (${\Sigma}_{\rm SFR}$) for our galaxy
sample. The lower and upper limits in the ${\Sigma}_{\rm SFR}$ are also
given together with the disk area used to normalize the SFRs and gas masses.}
\begin{tabular}{rrrrrr}
\hline
Galaxy   &   $\log{\Sigma}_{\rm g}$  & $\log{\Sigma}_{\rm SFR}$
& $\log{\Sigma}_{\rm SFR}^{\rm min}$ & $\log{\Sigma}_{\rm SFR}^{\rm max}$ &
Area \\ 
& $\displaystyle\frac{{\rm M}_{\odot}}{{\rm pc}^{2}}$ 
& $\displaystyle\frac{{\rm M}_{\odot}}{{\rm yr}\,{\rm kpc}^{2}}$ 
& $\displaystyle\frac{{\rm M}_{\odot}}{{\rm yr}\,{\rm kpc}^{2}}$ 
& $\displaystyle\frac{{\rm M}_{\odot}}{{\rm yr}\,{\rm kpc}^{2}}$ 
& ${\rm kpc}^2$ \\
\hline		   
NGC~891   & 1.09 & -2.30 & -2.54 & -2.07 & 687\\
NGC~5907  & 1.27 & -2.29 & -2.29 & -2.15 & 421\\
NGC~4013  & 1.40 & -1.85 & -1.85 & -1.63 &  88\\
UGC~1082  & 0.87 & -2.27 & -2.27 & -1.89 & 482\\
UGC~2048  & 0.70 & -2.80 & -3.17 & -2.60 & 855\\
\hline
\end{tabular}
\label{tab3}
\end{table}

\section{Summary}

\begin{itemize}
\item
This paper is the second part on a series of papers dedicated to 
the modeling of the SEDs
of disk galaxies. In Paper~I we described a new tool for the analysis of the UV
to the submm SED and applied this tool to the well known nearby edge-on galaxy
NGC~891. In the present paper we have extended the analysis from Paper~I to a
sample of four additional edge-on galaxies (NGC~5907, NGC~4013, UGC~1082,
UGC~2048) and proved that the solution obtained for NGC~891 is generally
valid. We have also shown that NGC~891 is not an exceptional system in terms of
its SFR density. 

\item In all the systems the dust is predominantly heated by the young stellar
population. The contribution of the old stellar population can account for at
most $40\%$. 

\item The SFRs derived from our ``standard model'' are consistent with the SFRs
derived from the more sophisticated ``two-dust-disk model'' and are also
consistent with the Kennicutt's (\cite{kennicutt}) Schmidt law for disk galaxies.

\end{itemize}

\begin{acknowledgements}
One of us (A.M.) thanks the Max Planck Institut f\"ur Kernphysik
for financial support and hospitality during his visit
to Heidelberg.
\end{acknowledgements}


\begin{thebibliography}{}


   \bibitem[1998]{alton}
   Alton, P. B., Bianchi, S., Rand, R. J., et al., 
   1998, ApJ, 507, L125 

   \bibitem[2000a]{bianchi1}
   Bianchi, S., Davies, J. I., Alton, P. B., Gerin, M., Casoli, F.,
   2000, A\&A, 353, L13

   \bibitem[2000b]{bianchi2}
   Bianchi, S.,  Davies, J. I., Alton, P. B.,
   2000, A\&A, 359, 65

   \bibitem[1995]{bottema}
   Bottema, R., 
   1995, A\&A, 295, 605

   \bibitem[1997]{bregman}
   Bregman, J. N., \& Houck, J. C., 
   1997, ApJ, 485, 159

   \bibitem[2001]{bruzual} 
   Bruzual, A. G., \& Charlot, S.,  
   2001, in preparation

   \bibitem[1999]{devrient} 
   Devriendt, J. E. G., Guiderdoni, B., Sadat, R., 
   1999, A\&A, 350, 381

   \bibitem[1997]{dumke}
   Dumke, M., Braine, J., Krause, M., et al., 
   1997, A\&A, 325, 124

   \bibitem [2000]{dunne}
   Dunne, L., Eales, S., Edmunds, M., et al., 
   2000, MNRAS, 315, 115

   \bibitem[2000]{efstathiou}
   Efstathiou, A., Rowan-Robinson, M., Siebenmorgen, R., 
   2000, MNRAS, 313, 375

   \bibitem[1992]{garcia}
   Garc\'{\i}a-Burillo, S., Gu\'elin, M., Cernicharo, J., Dahlem M., 
   1992, A\&A, 266, 21.

   \bibitem[1993]{giovanelli}
   Giovanelli, R., Haynes, M. P., 
   1993, AJ, 105, 1271

   \bibitem[1997]{gomez}
   Gomez de Castro, A. I., Garc\'{\i}a-Burillo, S., 
   1997, A\&A, 322, 381

   \bibitem[1993]{guelin}
   Guelin, M., Zylka, R., Mezger, P.G., et al., 
   1993, A\&A, 279, L37

   \bibitem[1998]{haas} 
   Haas, M., Lemke, D., Stickel, M., et al. 
   1998, A\&A, 338, L33

   \bibitem[1999]{hoopes}
   Hoopes, C. G., Walterbos, R. A. M., \& Rand, R. J., 
   1999, ApJ, 522, 669

   \bibitem[1989]{huchtmeier}
   Huchtmeier, W. K., Richter, O. G.,
   1989, A general catalog of HI observations of galaxies, New York, 
   Springer-Verlag 
 
   \bibitem[1999]{israel}
   Israel, F. P., ven der Werf, P. P., Tilanus, R. P. J., 
   1999, A\&A, 344, L83    

   \bibitem[1998]{kennicutt}
   Kennicutt, R. C. Jr., 
   1998, ApJ, 498, 541
 
   \bibitem[1998] {kuchinski}
   Kuchinski, L. E., Terndrup, D. M., Gordon, K. D., Witt, A. N., 
   1998, AJ, 115, 1438   

   \bibitem[1987]{kylafis}
   Kylafis, N. D., Bahcall, J. N., 
   1987, ApJ, 317, 637

   \bibitem[1993]{laor}
   Laor, A., Draine, B. T., 
   1993, ApJ, 402, 441

   \bibitem[1977]{mathis} 
   Mathis, J. S., Rumple, W., Nordsieck, K. H., 
   1977, ApJ, 217, 425

   \bibitem[2000]{misiriotis}
   Misiriotis, A., Kylafis, N. D., Papamastorakis, J., Xilouris, E. M.,
   2000, A\&A, 353, 117

   \bibitem[1990]{moshir}
   Moshir, M., et al.
   1990, IRAS Faint Source Catalogue.

   \bibitem[1996]{neininger} 
   Neininger, N., Gu\'elin, M., Garc\'{\i}a-Burillo, S., et al.,
   1996, A\&A, 310, 725

   \bibitem[1995]{ohta} 
   Ohta, K., Kodaira, K., 
   1995, PASJ, 47, 17

   \bibitem[2000]{pohlen}
   Pohlen, M., Dettmar, R.-J., Lutticke, R.,
   2000, A\&A, 357, 1

   \bibitem[2000]{popescu}
   Popescu, C. C., Misiriotis, A., Kylafis, N. D., Tuffs, R. J., Fischera, J.,
   2000, A\&A, 362, 138 (Paper I)

   \bibitem[1991]{rupen}
   Rupen, M. P., 
   1991, AJ, 102, 48

   \bibitem[1998]{silva}
   Silva, L., Granato, G .L., Bressan, A., Danese, L., 
   1998, ApJ, 509, 103

   \bibitem[1989]{soifer}
   Soifer, B. T., Boehmer, L., Neugebauer, G., Sanders, B.,
   1989, AJ, 98, 766

   \bibitem[2000]{stickel} 
   Stickel, M., Lemke, D., Klaas, U., et al. 
   2000, A\&A, 359, 865

   \bibitem[1997]{xilouris1}
   Xilouris, E. M., Kylafis, N. D., Papamastorakis, J., Paleologou,
   E. V., Haerendel, G.,
   1997, A\&A, 325, 135 (X97)

   \bibitem[1998]{xilouris2}
   Xilouris, E. M., Alton, P. B., Davies, J. I., et al.,
   1998, A\&A, 331, 894 (X98)

   \bibitem[1999]{xilouris3}
   Xilouris, E. M., Byun, Y. I., Kylafis, N. D., Paleologou, E. V., Papamastorakis, J.,
   1999, A\&A, 344, 868 (X99)

   \bibitem[1995]{xu1} 
   Xu, C., Buat, V., 
   1995, A\&A, 293, L65

   \bibitem[1996]{xu2} 
   Xu, C., Helou, G., 
   1996, ApJ, 456, 163

   \bibitem[1989]{young}
   Young, J. S., Xie, S., Kenney, J. D. P., Rice, W. L.,
   1989, ApJS, 70, 699

\end{thebibliography}
\end{document}